\documentclass[manuscript]{aastex61}
\usepackage{float,ltxtable,natbib}
\usepackage{multirow}
\usepackage{amssymb,rotating,amsmath}
\usepackage{url,enumerate}

\newcommand{\au}{\,\mathrm{au}}

\usepackage{color}

\shorttitle{Unconfirmed Near-Earth Objects}
\shortauthors{Vere\v{s} et al.}

\begin{document}

\title{Unconfirmed Near-Earth Objects}

\correspondingauthor{Peter Vere\v{s}}
\email{peter.veres@cfa.harvard.edu}

\author{Peter Vere\v{s}}
\affiliation{Minor Planet Center, Harvard-Smithsonian Center for Astrophysics, 60 Garden Street, Cambridge, MA 02138}

\author{Matthew J. Payne}
\affiliation{Minor Planet Center, Harvard-Smithsonian Center for Astrophysics, 60 Garden Street, Cambridge, MA 02138}

\author{Matthew J. Holman}
\affiliation{Minor Planet Center, Harvard-Smithsonian Center for Astrophysics, 60 Garden Street, Cambridge, MA 02138}

\author{Davide Farnocchia}
\affiliation{Jet Propulsion Laboratory, California Institute of Technology, 4800 Oak Grove Drive, Pasadena, CA, 91109}

\author{Gareth V. Williams}
\affiliation{Minor Planet Center, Harvard-Smithsonian Center for Astrophysics, 60 Garden Street, Cambridge, MA 02138}

\author{Sonia Keys}
\affiliation{Minor Planet Center, Harvard-Smithsonian Center for Astrophysics, 60 Garden Street, Cambridge, MA 02138}

\author{Ian Boardman}
\affiliation{Minor Planet Center, Harvard-Smithsonian Center for Astrophysics, 60 Garden Street, Cambridge, MA 02138}

\begin{abstract}

We studied the Near-Earth Asteroid (NEA) candidates posted on the Minor Planet Center's Near-Earth Object Confirmation Page (NEOCP)
between years 2013 and 2016. 
Out of more than 17,000 NEA candidates, 
while the majority became either new discoveries or were associated with previously known objects, 
about 11\% were unable to be followed-up or confirmed. 
We further demonstrate that of the unconfirmed candidates, $926\pm 50$ are likely to be NEAs, representing 18\% of discovered NEAs in that period. 
Only 11\% ($\sim93$) of the unconfirmed NEA candidates were large (having absolute magnitude $H<22$).
To identify the reasons why these NEAs were not recovered, we analyzed those from the most prolific asteroid surveys: Pan-STARRS, the Catalina Sky Survey, the Dark Energy Survey, and the Space Surveillance Telescope.
We examined the influence of plane-of-sky positions and rates of motion, brightnesses, submission delays, and computed absolute magnitudes, as well as correlations with the phase of the moon and seasonal effects.
We find that delayed submission of newly discovered NEA candidate to the NEOCP drove a large fraction of the unconfirmed NEA candidates. 
A high rate of motion was another significant contributing factor.
We suggest that prompt submission of suspected NEA discoveries and rapid response to fast moving targets and targets with fast growing ephemeris uncertainty would allow better coordination among dedicated follow-up observers, decrease the number of unconfirmed NEA candidates, and increase the discovery rate of NEAs.

\end{abstract}


\section{Introduction}

Near-Earth Objects (NEOs) are defined as having a perihelion distance smaller than 1.3$\au$ and represent a population of asteroids (NEA) and comets (NEC) with orbits near the Earth's orbit. The first NEA (433) Eros was discovered in 1898 using photographic plates. Since then, detection techniques have improved and the total number of known NEAs exceeded 18,000 in 2018. 
The discovery rate in the last 20 years has risen rapidly, primarily driven by the mandate of the US Congress to NASA in 1998 to discover 90\% of NEOs larger than 1\,km in the so-called Spaceguard Survey. 
In 2005, the US congressional mandate to NASA expanded the task to catalog 90\% of NEOs larger than 140\,m  within the next 15 years (George E. Brown, Jr. NEO Survey Act \footnote{Section 321 of the NASA Authorization Act of 2005 (Public Law No. 109-155)}). 
The motivations for discovering NEOs are diverse and include research related to the origin and evolution of the Solar System, planetary defense, planned spacecraft operations, 
and proposed commercial utilization (e.g. asteroid mining).

The discovery of an NEA is a process that requires distinguishing between known and unknown targets, and then following up (obtaining additional observations of) any unknown targets with the aim of extending the arc of observations and determining the object's orbit. 
The catalog of orbital elements of known NEAs, their size-frequency distribution,  as well as the region of sky visited by telescopes, all serve as inputs for deriving debiasied population models \citep{Bottke02,Grav11,Greenstreet12,Hinkle15,Harris15,Lilly17,Tricarico17}, NEA population statistics \citep{Jedicke98,Jedicke2015,Jedicke16} and Earth-impact frequency models  \citep{Brown02,Ivanov2001,Brown13}. 
However, one key element is usually missing from the analysis: NEA candidates that were observed, reported, but subsequently were not confirmed. For instance, the simulations rely on detection and non-detection of synthetic objects and its comparison to the catalog of discovered orbits, while the realistic processes of discovery and linking are far more complex. If systematic biases exist in the properties of unconfirmed NEAs, the population models derived from the surviving discoveries may also be biased.

In this work we describe the NEA discovery process and focus on the quantitative and qualitative reasons that cause NEA candidates to remain unconfirmed.

\section{Near-Earth Object Confirmation Page}
\label{s:NEOCP}

The Minor Planet Center (MPC) collects astrometry of asteroids and comets, calculates derived orbits, and publishes observations and orbits under the auspices of the International Astronomical Union (IAU). 
Observers submit astrometry to the MPC, where objects are either attributed to known orbits or are considered new. 
New discoveries that could be NEAs are posted by the MPC on the NEO confirmation page\footnote{\url{https://www.minorplanetcenter.net/iau/NEO/toconfirm_tabular.html}} \citep[NEOCP,][]{Marsden98}. 
A subsection of the NEOCP is dedicated to comet candidates.
Once additional observations are collected and an orbit calculated, the object is added to the catalog of known objects.

The purpose of the Near-Earth Object Confirmation Page (NEOCP) is to provide real-time publication of NEA candidates and facilitate the rapid follow-up and initial orbit determination of NEAs. Even though a short one-night arc is generally insufficient for orbit determination, tools like the {\it digest2} code\footnote{Keys et al., {\it in prep.}} can provide a statistical estimate as to whether the detections belong to an NEA
\footnote{Source code at \url{https://bitbucket.org/mpcdev/digest2/src}}. 

The {\it digest2} code explores a range of bound orbits (i.e. orbits with eccentricity less than one) in the so-called admissible region \citep{Milani04, Milani05, Farnocchia15}, then compares the orbits to a prior population model based on the known catalog of orbits, and returns a quasi-probability score, $D_2$, in a range $0\leq~D_2\leq100$.
Submissions to the MPC with the NEOCP keyword and $D_2\geq65$ are placed on the NEOCP, where it becomes publicly available to other observers who can provide additional follow-up observations.

Submissions to the NEOCP have a number of possible fates:

\begin{enumerate}[(i)]
\item They may be associated with the orbit of an already known object (which may or may not be an NEA): this is called {\it attribution}; 
\item If additional detections (from dedicated follow-up and/or serendipitous survey observations) can be associated with the initial submission, and the observational arc becomes long enough to determine its orbit, a new {\it discovery}  is announced (again,  either an NEA or a non-NEA); 
\item The submission can be {\it retracted} by observers as being erroneous or {\it not a minor planet}; 
\item The object may remain unconfirmed, with no additional observations and is removed from NEOCP.
\item The object may be reported as a {\it comet}.
\end{enumerate}

The current MPC criteria for the {\it successful} removal of NEOCP candidates are based on the absolute magnitude \citep[$H$,][]{Bowell89}, the length of the observational arc, and the number of tracklets\footnote{A tracklet is a set of detections taken from a single site within a short time-frame, typically an hour}. 
These criteria are presented for the sake of completeness in Table~\ref{neocp} of our Appendix. 
When an NEA or an unusual object ($a>5.8\au$ or $e>0.5$) is discovered as the result of being posted on the NEOCP, the MPC announces the object through a Minor Planet Electronic Circular (MPEC). 
Improved astrometric accuracy and the availability of high quality star catalogs in recent years are such that arc-lengths of a few days are typically sufficient for the successful determination of an orbit and  the removal of a candidate from the NEOCP. 
For instance, an NEA with $H=22$, an observational arc of 2 days and containing 4 distinct tracklets is sufficient to be announced through an MPEC.

If no follow-up observations are made and the object's sky-plane uncertainty grows to such an extent that practical follow-up becomes impossible, e.g. the uncertainty region becomes an order of magnitude larger than the field of view of a follow-up telescope, the candidate is removed from the NEOCP. Such a candidate is moved to the MPC's ``isolated tracklet file'' (ITF\footnote{\url{https://www.minorplanetcenter.net/iau/ECS/MPCAT-OBS/MPCAT-OBS.html}}). We note that the ITF contains all submissions (5.5 million tracklets as of February 2018) that have not yet been attributed to any known orbit (including submissions not sent to the NEOCP). 
The ITF is regularly processed by the MPC, and linkages leading to new orbits or attributions to known orbits are regularly found by the MPC and the external users \citep[e.g.][]{Weryk17}. After a successful linkage with an orbit, a tracklet is removed from ITF.

In this paper we are interested in the NEA candidates that were posted on NEOCP but were removed as unconfirmed. 
Identifying the underlying issues that prevent their confirmation is complex.  There are a number of possible biases influencing the decisions of NEO surveys and follow-up observers.  The key driver for NEO observers is to deliver discoveries, and decisions must be made as to how to deploy limited resources in an effort to maximize the discovery rate.  It is plausible that ``challenging'' NEA candidates that are either too faint, moving too fast, or have significant positional uncertainties, are given a lower priority for follow-up, which in turn may lead to systematic biases in the properties of the NEAs that remain unconfirmed. We examine a sample of NEOCP candidates and study the reasons why they remain unconfirmed, including both quantitative and qualitative aspects.

\section{Data}
\label{s:data}

Our sample of NEA candidates contains 3.5 years of NEOCP data from 2013 to 2016, with the exception of the time period between January and July 2014.  In total, 17,030 candidates appeared on the NEOCP in the studied time frame.
Table~\ref{neocp_table} provides a breakdown of the fate of these candidates. 
The most common outcome was that a candidate resulted in either the ``discovery'' or ``attribution'' of a non-NEA object ($50\%$ of cases), while in $31\%$ of cases the NEOCP candidate turned out to be an NEA, but in approximately $11\%$ of cases the NEA candidate was removed from NEOCP as being ``initially unconfirmed''.

The majority of the remainder were tracklets that were subsequently attributed to previously known NEAs. 
A small fraction ($2\%$) were retracted by observers as objects that were not real (``retracted'') or subsequently identified by MPC as artificial satellites or space debris (``not a minor planet''). The NEOCP also provides information on new ``comet'' candidates. In this work we exclude comets, objects that are not a minor planet, and retracted false objects from further analysis, and focus on confirmed asteroids and unconfirmed NEOCP candidates (16,280).
\begin{table*}[ht!]
\small
\begin{center}
\caption{\bf The classification of all NEOCP candidates between 2013 and 2016 (see Section\ref{s:NEOCP} and \ref{s:data} ). Out of 1,909 NEA candidates that were initially unconfirmed, 315 were later confirmed (see Table \ref{lostitf}). In the rest of this work we will consider ``unconfirmed'' NEA candidate only if it remains presently unknown. 
We will further analyze only the asteroid-like NEA candidates (16,280) and focus on those with $D_2=100$ (see Section \ref{s:data:digest} for detailed discussion). }
\begin{tabular}{l|c|c||c}
\hline
       &   Number   &       & Number with    \\
 Classification  &  & (\%)  & $D_2=100$    \\
\hline\hline
Initially Unconfirmed&1,909&11&915\\
NEA Discovery &5,117&31&3,768\\
NEA Attribution&708&4&492\\
Non-NEA (Discovery \& Attribution) &8,546&50&184\\
\hline
Comet&231&1&67\\
Not a minor planet&109&1&92\\
Retracted&410&2&230\\
\hline\hline
Summary&17,030&100&5,748\\
\hline
\end{tabular}
\label{neocp_table}
\end{center}
\end{table*}

Even though the orbit of an NEA is officially published through an MPEC, and the NEA candidate then removed from the NEOCP, it is up to the individual observer to decide whether or not they keep observing an object. 
If no follow-up data are collected, the observation arc will remain short and object's orbit will be too poorly determined to allow accurate future predictions of its position.
 
\begin{figure}[ht!]
\center
\includegraphics[width=0.48\textwidth]{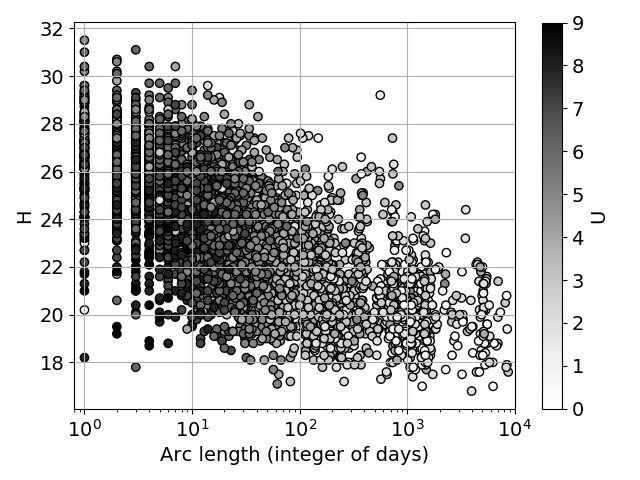}
\includegraphics[width=0.48\textwidth]{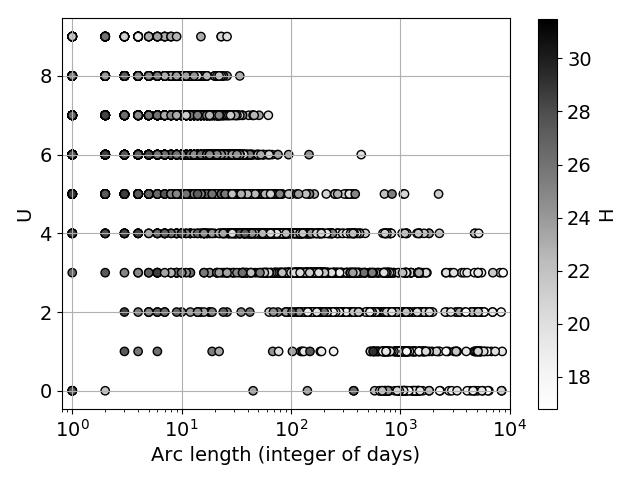}

\caption{Scatter plot of orbit arc and absolute magnitude (left) and uncertainty parameter U (right) of NEA discoveries between years 2013-2016. Large U means large orbit uncertainty.}
\label{MPEC_U}
\end{figure}

{\bf Figure~\ref{MPEC_U} plots the absolute magnitude, $H$, and the orbital uncertainty $U$
\footnote{\url{https://minorplanetcenter.net/iau/info/UValue.html}}}
\citep{Marsden78} as functions of arc length.
This demonstrates that short arcs are typical for the smallest objects (large $H$) that are only visible for a short time and their orbit quality, denoted by the uncertainty parameter U, is usually lower than that of larger objects.

We note that of the $1,909$ NEOCP candidates that were initially removed from the NEOCP as being unconfirmed, 315 ($16\%$) were later attributed  to previously or later designated asteroids (see Table~\ref{neocp_table}, ~\ref{lostitf}). 
Out of these 315 initially unconfirmed NEOCP candidates that were later linked, the majority ($84\%$) were non-NEAs and only $16\%$ were NEAs (Table ~\ref{lost_itf_fraction}).
In the remainder of the paper, we shall focus on the $1,594$ NEOCP candidates that remained unconfirmed as of February 2018, and will refer to these generally as being the ``unconfirmed'' candidates.

\begin{figure}[ht!]
\center
\includegraphics[height=0.42\textwidth]{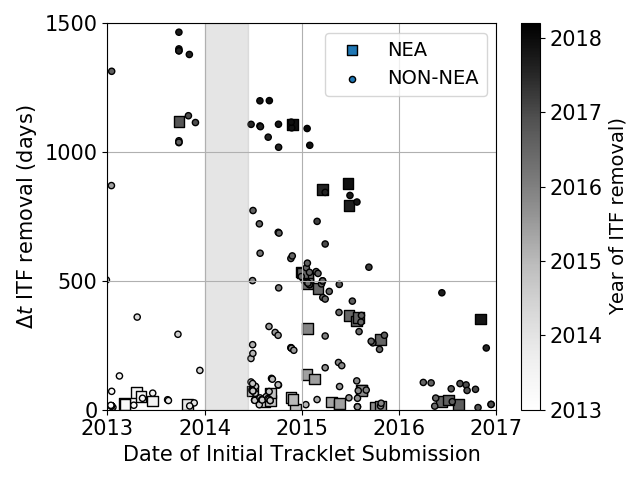}
\caption{Unconfirmed NEA candidates that were later attributed or linked. The grey area represents missing data. It takes days to years to attribute unconfirmed NEOCP candidate. The apparent diagonal clustering is a consequence of objects and attributed at the next apparition.}
\label{itf}
\end{figure}

The plot in Figure~\ref{itf} shows how long the unconfirmed NEOCP candidates remained in the ITF before they were linked. 

We note that the ``Date of Initial Tracklet Submission'' plotted in Figure \ref{itf} corresponds to the date of submission of the tracklet that went into the ITF and the $\Delta\,t$ plotted on the y-axis corresponds to the actually amount of time it spent in the ITF. 
This means that the plotted submission date may well differ from the official ``Discovery Date'' of the object with which the tracklet is subsequently attributed, due to the system of priorities used by the MPC to assign official discovery credit\footnote{https://www.minorplanetcenter.net/mpec/K10/K10U20.html}.

It is clear that objects linked near the end of our study's time frame have had less time to be rediscovered by surveys than those earlier.
To account for this, we extrapolate the rate of recovery from earlier years, and thus estimate that a small number of discoveries from 2015 and 2016 that are currently in the ITF will likely get linked and attributed in the coming months and years. 
We estimate that by the end of 2018 this will reduce the current unconfirmed numbers (1,594) down to $1,524\pm7$. 
It is apparent that the ITF contains many unattributed objects and its linking remains one of the MPC's priorities.

\begin{table*}[ht!]
\small
\begin{center}
\caption{Initially unconfirmed and later attributed NEOCP candidates. In the rest of this work we will consider only the ``Currently Unconfirmed'', and we will focus on $D_2=100$ candidates (see Section \ref{s:data:digest} for further discussion).}
\begin{tabular}{p{9cm}|c|c}
\hline
	&All	&$D_2=100$\\
\hline
Initially Unconfirmed	&1,909&	915\\
Initially-unconfirmed-but-subsequently-attributed&	315	&95\\
\bf{Currently Unconfirmed}	&\bf{1,594}	&\bf{820}\\
\hline
\end{tabular}
\label{lostitf}
\end{center}
\end{table*}

\begin{table*}[ht!]
\small
\begin{center}
\caption{Unconfirmed NEA candidate attributed to NEAs and non-NEAs between 2013-2016.}
\begin{tabular}{l|c|c}
\hline

	&All&	$D_2=100$\\
\hline
NEA&	16\%&	36\%\\
Non-NEA	&84\%&	64\%\\
\hline
\end{tabular}
\label{lost_itf_fraction}
\end{center}
\end{table*}

\section{ {\sc Digest2} Score and estimation of unconfirmed NEA count}
\label{s:data:digest}

\begin{figure*}[ht!]
\center
\includegraphics[width=0.49\textwidth]{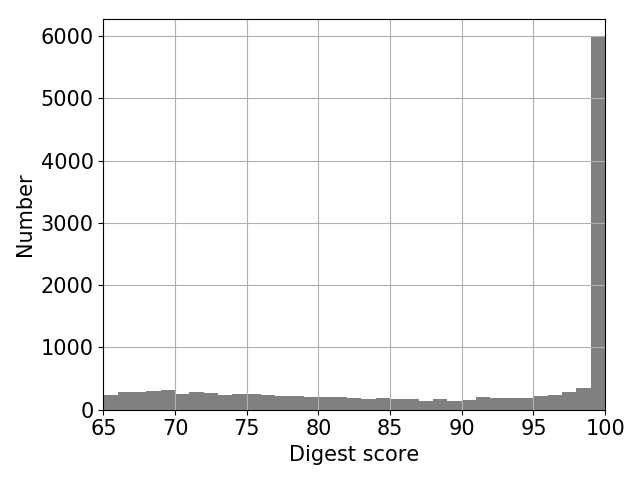}
\includegraphics[width=0.49\textwidth]{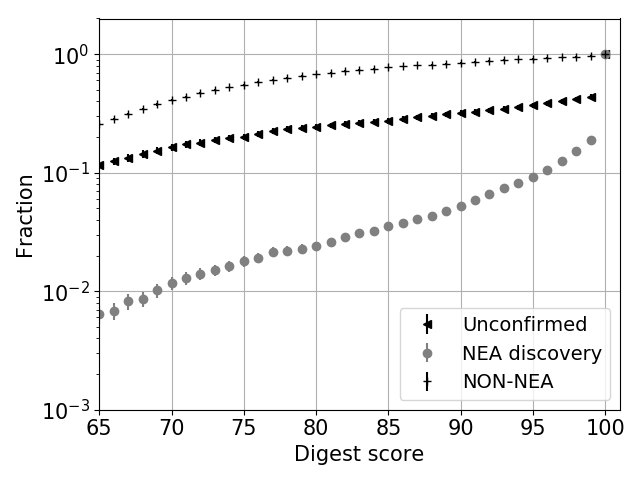}
\caption{%
{\bf Left:} Histogram of  the {\it digest2} score, $D_2$ for submitted NEOCP candidates and {\bf Right:} a normalized cumulative distribution of the {\it digest2} score, $D_2$, for unconfirmed NEA candidates, discovered NEAs, and discovered non-NEAs.
}\label{digestscore}
\end{figure*}

\begin{figure*}[ht!]
\center
\includegraphics[width=0.6\textwidth]{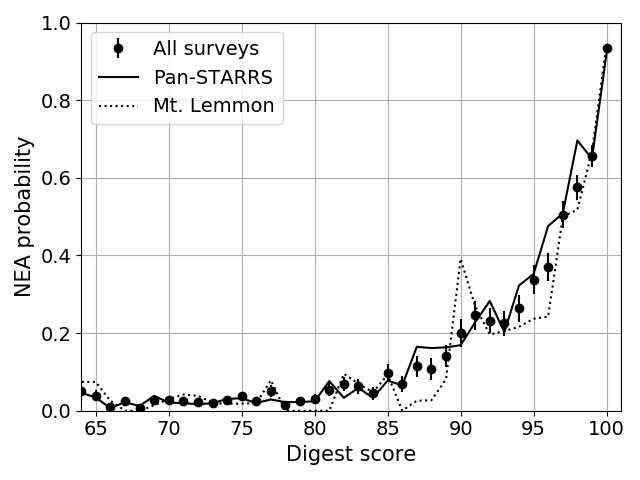}
\caption{%
The probability of an NEOCP submission being an NEA as a function of {\it digest2} score. }
\label{digestscore1}
\end{figure*}

The {\it digest2} score, $D_2$, is a classifier that distinguishes NEAs from non-NEAs, particularly for single tracklets and very short arcs where orbit determination provides a wide range of solutions. 
We used a population model \citep{Bottke02} to set a score for each possible orbit. Moreover, the score is scaled for the expected fraction of undiscovered population in the corresponding orbit class. 
With the goal of understanding the likely orbit class of the unconfirmed candidates, in this and subsequent sections we compare the $D_2$ score for unconfirmed objects, with the $D_2$ score for the first tracklet of the {\it discovered} and identified objects. 
The statistics of the $D_2$ score for submissions to the NEOCP is shown on Figure~\ref{digestscore}. About 33\% of all NEOCP candidates (and $52\%$  of {\it unconfirmed} candidates) have $D_2=100$. 
The behavior is strikingly different between the non-NEA and NEAs: 90\% of NEAs had $D_2>90$, whereas 90\% of non-NEAs had $D_2<95$. 

We note that one should not consider $D_2$ as a binary discriminator, but rather as a statistical tool. A low-digest2 NEOCP candidate still has a low, but non-zero probability of being an NEA, while a high-scoring NEOCP candidate may still be a non-NEA (albeit with low probability). 
The right-hand side of Figure~\ref{digestscore} shows that unconfirmed NEA candidates generally have large values of $D_2$, which means that their plane-of-sky motion resembles that of NEAs.

We would like to estimate what fraction of unconfirmed candidates might be NEAs.
To do this we begin by calculating an approximate ``NEA-probability'' for the successfully identified NEOCP candidates as a function of $D_2$ by dividing the actual number of NEAs (sum of discovered and attributed NEAs) by the number of NEOCP candidates for each binned value of $D_2$.  We assume that the same probability extrapolates to unconfirmed objects. 

The resulting probability is displayed on the right of Figure~\ref{digestscore1}.  At $D_2=100$, an NEOCP object has a $93\pm1\%$ probability\footnote{We use the central limit theorem for error estimation.} of being an NEA. 
The probability drops quickly: there is only a $37\pm7\%$ probability that an object is an NEA at $D_2=95$. 
By multiplying the NEA-probability by the number of unconfirmed NEA candidates per $D_2$ bin, we obtain an estimate of the number of unconfirmed candidates that are actually NEAs, assuming the same distribution  of $D_2$ for unconfirmed candidates.

We estimate that in the studied time period, $926\pm50$ unconfirmed candidates were NEAs. 
In comparison to the 5,117 discovered NEAs, if all the unconfirmed candidates had been recovered, the number of discoveries would have increased by 18\%. 
Out of the estimated $926\pm50$ NEAs coming from unconfirmed candidates, the majority, $828\pm11$, had $D_2=100$.

In this study we focus on the $D_2=100$ unconfirmed NEO candidates in detail because they are the most likely to be NEAs.

\section{Characterization of unconfirmed NEA candidates}

\subsection{Surveys and magnitudes}  

\begin{figure}[ht!]
\center
\includegraphics[width=0.49\textwidth]{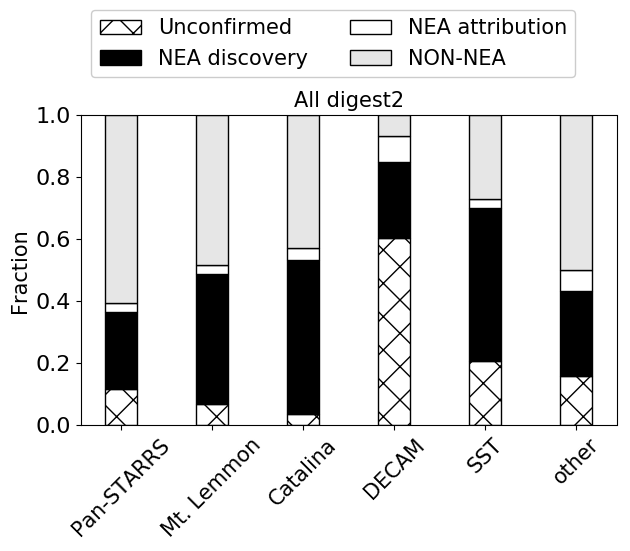}
\includegraphics[width=0.49\textwidth]{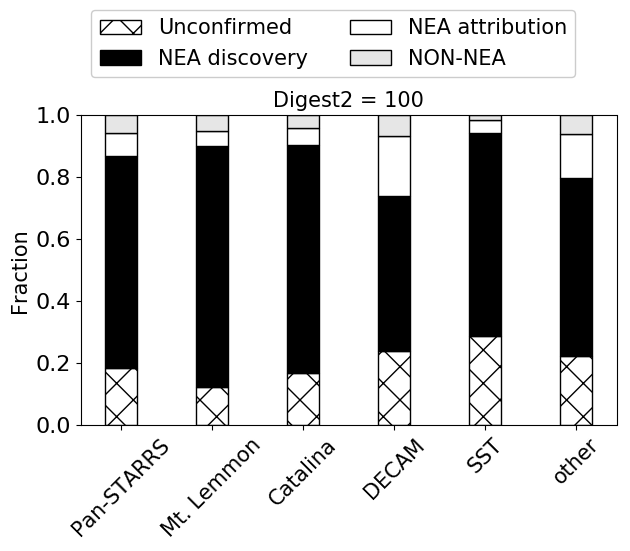}
\caption{Classification of NEOCP candidates and the fractional distribution for the most prolific surveys (left) and for $D_2=100$ NEOCP candidates (right). 
The loss-to-discovery ratio varies greatly among surveys. 
A large fraction of $D_2=100$ NEOCP candidates turn into NEA discoveries.}
\label{class_fraction}
\end{figure}

Most NEOCP candidates are submitted by large surveys, such as Pan-STARRS, Catalina Sky Survey (Mt. Lemmon and Catalina telescopes), Space Surveillance Telescope (SST), and the Dark Energy Camera (DECAM). 
Details of the most prolific NEO surveys are provided in Table~\ref{tsurveys2}.  Figure~\ref{class_fraction} shows that the ratio of unconfirmed to all NEOCP candidates varies greatly from low (Mt. Lemmon, Catalina Sky Survey, SST: 4-9\%) , through medium (Pan-STARRS, 9-17\%), to large (DECAM, 59-62\%). The overall NEOCP loss rate is $10\%$, while for $D_2=100$ NEOCP candidates it is $15\%$.

Surveys  differ in their capabilities, scope, and coverage. 
The Catalina Sky Survey (CSS) and Pan-STARRS operate on a nightly basis and CSS have their own follow-up facilities.
SST submits NEOCP candidates with a significant time delay (See Section \ref{s:delay} and Figure~\ref{delay_followup} for more detail).
DECAM only operates on few nights per year and reaches very faint magnitudes, producing a record number of NEOCP candidates in a short time. However, DECAM does not have follow-up specific capabilities and therefore a large fraction of their candidates do not get confirmed.

\begin{table*}[ht!]
\tiny
\begin{center}
\caption{Statistics of discovered NEAs and unconfirmed NEOCP candidates for all and $D_2=100$ candidates.
}
\begin{tabular}{|l|c|cc|c||cc|c|}
\hline
  \multicolumn{2}{|c|}{} & \multicolumn{3}{c||}{All NEOCP Candidates}& \multicolumn{3}{c|}{NEOCP Candidates with $D_2$=100}  \\
 \hline
Observatory  &  MPC  &   Discovered NEA  &  Unconfirmed  &  NEOCP  &  Discovered NEA   &  Unconfirmed  &  NEOCP\\
\hline
Pan-STARRS   &   F51   &  2225  &  835  &  8974  &  1426  &  346  &  2096\\
Mt. Lemmon   &   G96   &  1558  &  217  &  3705  &  1251  &  149  &  1570\\
Catalina   &  703  &  747  &  49  &  1508  &  686  &  34  &  812\\
SST   &   G45   &  116  &  14  &  243  &  91  &  10  &  143\\
DECAM   &   W84   &  143  &  336  &  572  &  77  &  196  &  315\\
Other   &      &  328  &  143  &  1278  &  237  &  85  &  423\\
\hline
Total  &      &  5117  &  1594  &  16280  &  3768  &  820  &  5359\\
\hline
\end{tabular}
\label{tsurveys2}
\end{center}
\end{table*}

\begin{figure}[ht!]
\center
\includegraphics[width=0.49\textwidth]{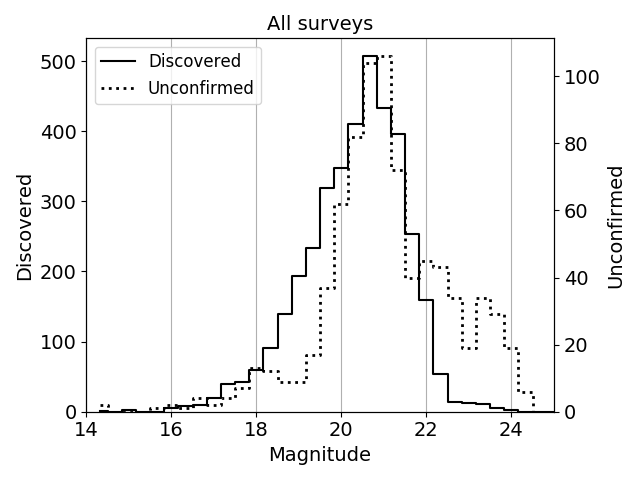}
\includegraphics[width=0.49\textwidth]{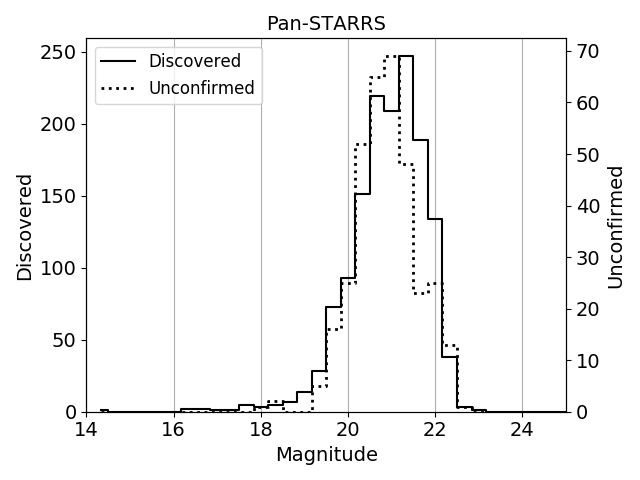}
\includegraphics[width=0.49\textwidth]{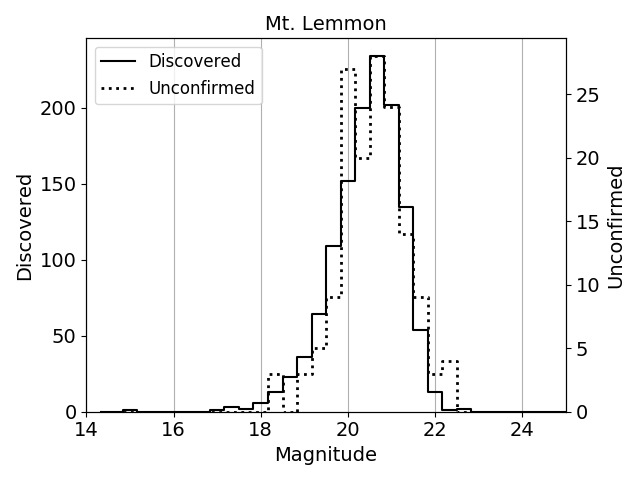}
\includegraphics[width=0.49\textwidth]{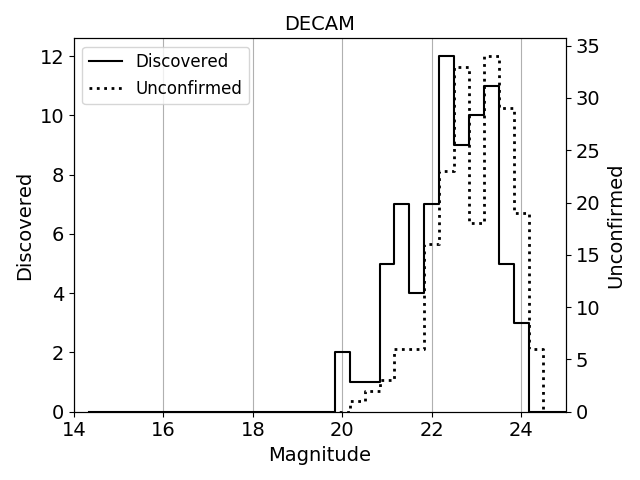}
\caption{V-band magnitude of discovered and unconfirmed NEAs by surveys, $D_2=100$. The magnitudes of the unconfirmed and discovered NEAs are similar for individual survyes.}
\label{magnitude}
\end{figure}

The size of the telescope, the exposure time,  and the image quality (seeing) determine the limiting magnitude of each exposure, and because populations of asteroids have a power law dependence on size \citep[e.g.][]{Bottke02}, larger telescopes detect more asteroids, the majority of which are faint. 
In addition, faint targets require larger telescopes or longer exposure times to be detected. 
Figure~\ref{magnitude} shows the V-band magnitude of unconfirmed and discovered NEA candidates ($D_2=100$) for all surveys, Pan-STARRS, Mt. Lemmon and DECAM. 
The conversion to V-band was done using transformations listed in Table~\ref{vband} of our Appendix. 
The histogram and mean-magnitudes of unconfirmed and discovered NEAs are similar in most cases. 
Even though the histogram for all magnitudes displays a relative excess of fainter magnitudes for unconfirmed NEAs, this is due to the contribution of DECAM and their extreme ratio of unconfirmed-to-discovered NEAs. 
Therefore, when analyzed on a survey-by-survey basis, we do not see a bias in the ratio of unconfirmed-to-discovered NEAs as a function of apparent magnitude.

Apparent magnitudes are used to determine the absolute magnitude and the asteroid's size. The absolute magnitude $H$ is defined as the brightness of an object at $1\,au$ from both the Earth and Sun and at a zero phase angle. In a photometric Johnson's V-band, the absolute magnitude is a key parameter for Solar System objects and is used to compute object's brightness within its ephemeris. 
With a known albedo, $H$ can be converted into an object's diameter. 
Because direct measurements of NEA diameters are difficult and only a negligible fraction of NEAs have been either visited by spacecraft, imaged by radar or observed in infrared, $H$ is primarily used for population modeling and size-frequency distribution estimates, e.g. \citep{Bottke02,Grav11,Greenstreet12,Hinkle15,Harris15,Lilly17,Tricarico17}.

For objects with well-defined orbits, $H$ can be calculated from the observed magnitude using a relationship such as the phase function \citep{Bowell89,Mui10}, because the distance to the object at the time of observation is known.
For unconfirmed NEOCP candidates, no single precise orbit is known, and a precise $H$ cannot be obtained. Instead, as described in our introduction, its short-arc orbit can be constrained to lie in the ``admissible region'' \citep{Milani04, Farnocchia15}, and from this, an approximate value calculated for $H$.

We used the {\it emcee} code \citep{Foreman2013} to select the orbits from within the admissible region that fit all detections in the submitted tracklet for $D_2=100$ NEA candidates.  We adopt a uniform prior for the points in the admissible region.
We generated initial conditions for the {\it emcee} code by randomly selecting Keplerian orbits from the admissible region for each submission, and we then weighted each observation by the appropriate astrometric uncertainty from \citep{Veres17}. 
We then used each of the output ``best-fit orbits'' to compute the range of possible $H$, using the IAU Two-Parameter Magnitude System for Asteroids by \citep{Bowell89},  assuming a slope parameter $G=0.15$.

\begin{figure}[ht!]
\center
\includegraphics[width=0.5\textwidth]{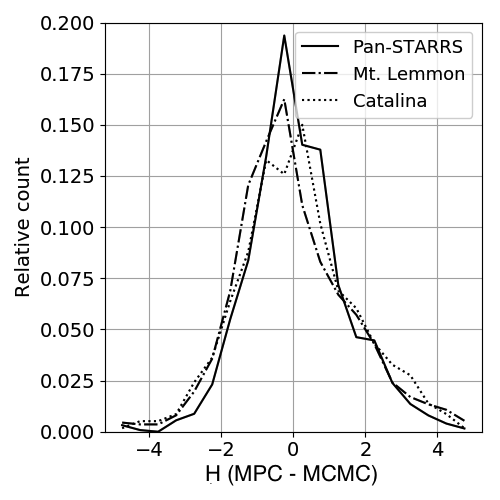}
\caption{Histograms of difference between known $H$ derived by MPC for entire orbit and $H$ derived by MCMC for a discovery tracklet for NEAs discovered between 2013-2016. Mean near zero means that MCMC estimated $H$ is near the true $H$.}
\label{deltaH}
\end{figure}

To verify this approach, we applied  {\it emcee} to the discovery tracklets from NEAs discovered between 2013-2016, thus, simulating analysis of unconfirmed NEO candidates that are mostly a single tracklet.
Since their orbits are known, $H$ is also determined. We compare the known $H$ for these NEAs with the value we derived from our MCMC approach and
confirmed that the mean difference is near zero (Figure~\ref{deltaH}). The uncertainty on the derived $H$ is large, with an average $1.6\,mag$. This value is consistent with the standard deviation of the difference between known and computed $H$, which is is $\sigma_{\Delta H}=1.9$ $mag$. We note that a large uncertainty is expected due to the large volume of orbital distances allowed within the admissible region.

\begin{figure}[ht!]
\center
\includegraphics[width=0.7\textwidth]{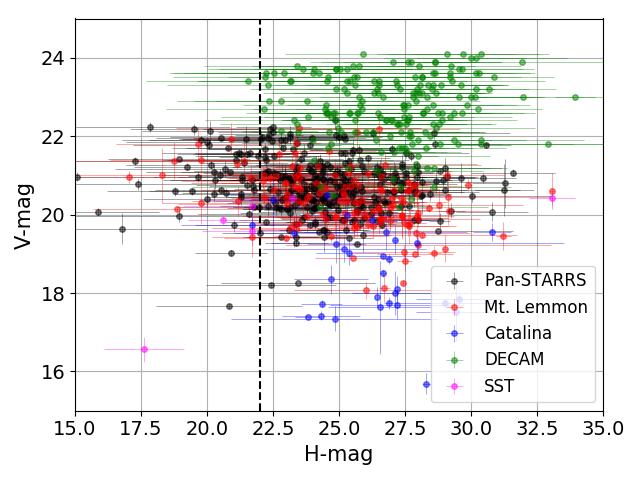}
\caption{Absolute vs. apparent magnitude in V-band of unconfirmed NEOCP candidates with $D_2=100$. Plots shows that DECAM is finding small NEAs with large $H$, which are the faintest. Larger objects with $H<22$ are found by Pan-STARRS, Mt. Lemmon and Catalina.}
\label{mag_scatter}
\end{figure}

Tables~\ref{Htable},~\ref{Htable1} and Figure~\ref{mag_scatter} analyze the apparent and absolute magnitudes of the unconfirmed NEOCP candidates and NEO discoveries coming from the NEOCP, focusing on the most prolific surveys.

\begin{table*}[ht!]
\tiny
\begin{center}
\caption{Mean apparent and absolute magnitude with RMS of unconfirmed NEA candidates with $D_2=100$. The last four columns divide the objects into different absolute magnitude ranges.} \begin{tabular}{l |c|c| c|| c| c| c| c}
\hline			
Observatory	& $\bar{V}$& $\bar{H}$ &Unconfirmed NEAs & $H<22$ & $22<H<25$ & $25<H<28 $& $H>28$\\
\hline
Pan-STARRS &$20.7\pm0.9$ &$24.4\pm2.6$  &346 & 56 & 148 & 118 & 24\\
Mt. Lemmon & $20.4\pm0.8$& $25.2\pm2.5$&149 & 13 & 48 & 76 & 12\\
Catalina & $18.6\pm1.2$&$26.2\pm2.0$  &34 & 1 & 9 & 19 & 5\\
DECAM & $22.7\pm0.8$ &$27.0\pm2.3$  &195 & 1 & 44 & 84 & 66\\
SST & $19.8\pm.1.2$ & $22.3\pm4.1$   &10 & 7 & 2 & 0 & 1\\
Other & $18.8\pm2.9$ & $25.9\pm5.7$ &86 & 15 & 15 & 24 & 32\\
\hline
Total & $20.8\pm1.8$& $25.3\pm3.2$&820 & 93 & 266 & 321 & 140\\
\hline
\end{tabular}
\label{Htable}
\end{center}
\end{table*}	

\begin{table*}[ht!]
\tiny
\begin{center}
\caption{Mean apparent and absolute magnitude with RMS of discovered NEAs. The last four columns divide the objects into different absolute magnitude ranges.
}

\begin{tabular}{l |c|c| c|| c| c| c| c}
\hline			
Observatory&	$\bar{V} $&$\bar{H}$ &Discovered NEAs & $H<22$ & $22<H<25$ & $25<H<28 $& $H>28$\\
\hline
Pan-STARRS & $20.9\pm0.8 $& $22.6\pm2.4$  & 2225 & 900 & 918 & 328 & 16\\
Mt. Lemmon  & $20.3\pm0.8 $ & $24.2\pm2.5$ & 1558 & 310 & 596 & 537 & 71\\
Catalina  & $18.9\pm0.8 $ &$23.8\pm2.7$ & 747 & 191 & 270 & 238 & 35\\
DECAM  & $ 22.4\pm0.9$ &$24.8\pm2.2$ & 143 & 18 & 45 & 70 & 7\\
SST  & $ 20.0\pm0.6$ &$22.4\pm2.1$ & 116 & 53 & 48 & 13 & 0\\
Other & $19.2\pm2.3$& $21.7\pm2.3$ & 328 & 198 & 98 & 27 & 1\\
\hline
Total &  $ 19.0\pm1.0$ &$23.3\pm2.6$ & 5117 & 1670 & 1975 & 1213 & 130\\
\hline
\end{tabular}
\label{Htable1}
\end{center}
\end{table*}

Objects%
\footnote{Objects with $H=22$ correspond to an approximate diameter of 140 meters, for the definition of a potential hazardous asteroid (PHA).} 
with $H<22$ are less frequent among unconfirmed NEA candidates than they are in the successfully recovered population (Table~\ref{Htable}). 
The majority of  $H<22$ unconfirmed NEA candidates were reported by Pan-STARRS.
The mean difference between the absolute magnitude of the discovered and unconfirmed NEA candidates is $\Delta H=2.0\,mag$: unconfirmed NEA candidates are smaller than discovered objects. 

Interestingly, the difference in the mean $H$ between unconfirmed and discovered NEA candidates differs across surveys. For instance Pan-STARRS has $\Delta H=1.8\,mag$, DECAM $\Delta H=2.2\,mag$  and Mt. Lemmon $\Delta H=1.0\,mag$. 
{\bf The small difference between the unconfirmed and the discovered NEAs demonstrates that Mt. Lemmon more efficiently retrieves NEA candidates. }
In contrast, Pan-STARRS loses proportionally more small NEA candidates, while DECAM has the largest $\Delta H$ due to their survey priorities, lack of follow-up and deep limiting magnitude. 
We note that DECAM contributed the largest number of the smallest unconfirmed NEAs (Table~\ref{Htable}): of the population with $H>28$, DECAM contributed more than half of them.

Approximately $11\%$ of the 820 unconfirmed NEA candidates with $D_2=100$  have $H<22$. 
Therefore, when we extrapolate this to the estimated {\it total} $926\pm50$ unconfirmed NEAs, we estimate that $102\pm6$ had $H<22$.

\subsection{Time delay}
\label{s:delay}

\begin{figure}[ht!]
\center
\includegraphics[width=0.45\textwidth]{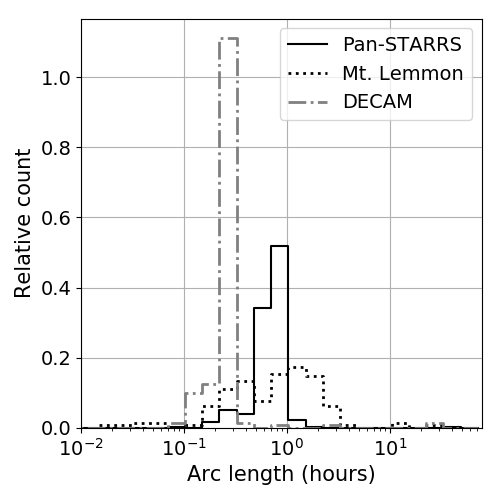}
\includegraphics[width=0.45\textwidth]{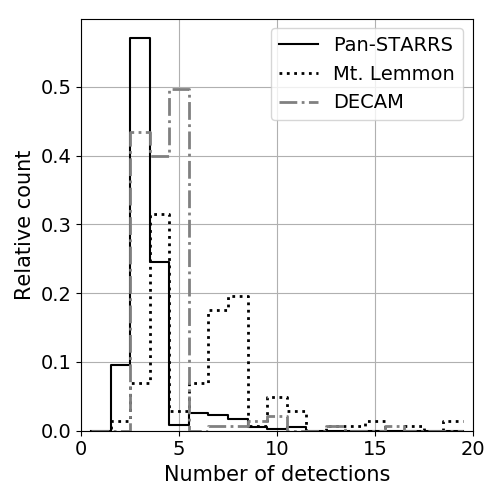}
\caption{Histogram of arc length (left) and number of detections (right) for unconfirmed candidate for Pan-STARRS, Mt. Lemmon and DECAM. Unconfirmed candidates are predominantly short, single-tracklet submissions, with no subsequent follow-up.}
\label{stats}
\end{figure}

Rapid follow-up of NEOCP candidates helps early arc extension, quick preliminary orbit determination and eases follow-up attempts on the second night. Figure \ref{stats} shows the distribution of observational arcs and number of detections for our set of unconfirmed NEA candidates. As expected, the observational arcs of unconfirmed candidates are short, typically having 2-6 detections, that only span a few hours and represent a single tracklet. In fact, only 17 out of 1,594 unconfirmed NEA candidates had any follow-up observations at all. The majority only had a single initial tracklet submission, without any reported follow-up.

Minor planet observations are not reported to MPC immediately.
The minimum time delay is typically equal to the time needed to acquire a minimum number of detections for a tracklet (typically four or more) plus the time to process the data. 
Because the initial tracklet cannot be represented by a single orbit, its positional uncertainty grows rapidly within a few hours of observation: Late submission can reduce the chances of collecting follow-up data due to this growing uncertainty, as well as to the day/night limitations and availability of follow-up telescopes around the world. 

\begin{figure}[htp!]
\center
\includegraphics[width=0.49\textwidth]{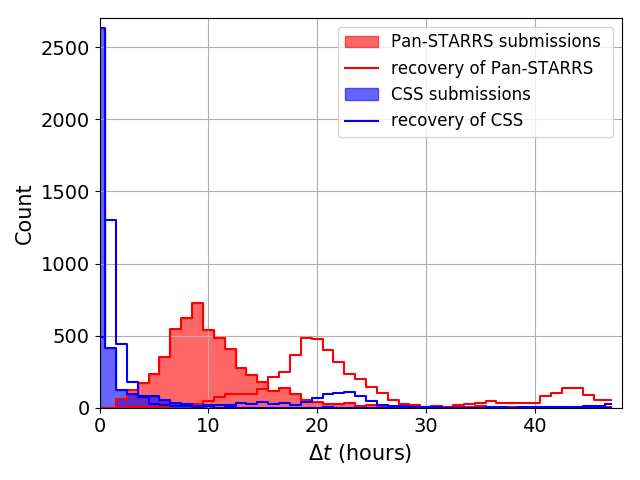}
\includegraphics[width=0.49\textwidth]{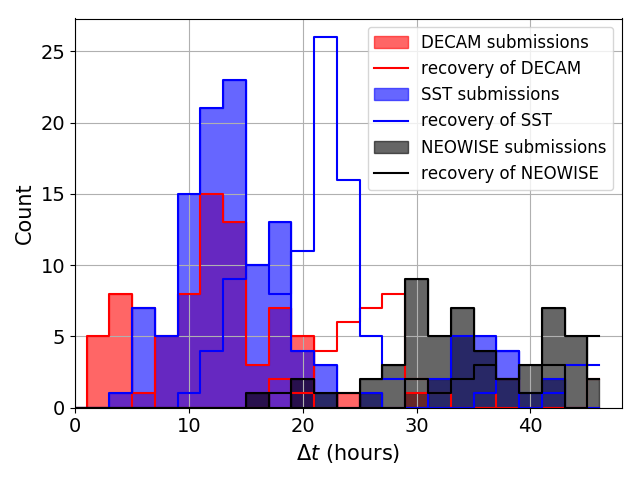}
\caption{Histograms of submission and follow-up time delay of confirmed NEOCP candidates for Pan-STARRS, Catalina Sky Survey (Catalina and Mt. Lemmon) (left) and DECAM, SST and NEOWISE (right). Submission delay varies greatly among surveys and large submission delay implies even larger recovery delay.}
\label{delay_followup}
\end{figure}

\begin{figure}[htp!]
\center
\includegraphics[width=0.75\textwidth]{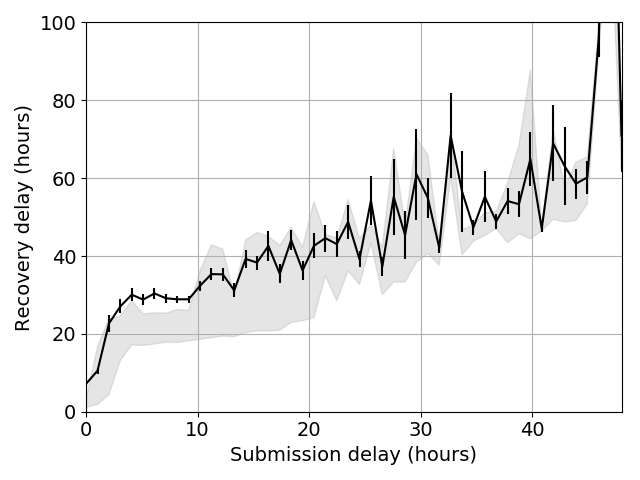}
\caption{Mean recovery delay as a function of the submission time delay for confirmed NEOCP candidates. 25-75 percentile is depicted by grey area. Delayed submission increases the recovery time.}
\label{delay_followup1}
\end{figure}

Figures~\ref{delay_followup} and ~\ref{delay_followup1} show the submission and recovery time delays for major asteroid surveys for {\it confirmed} NEOCP candidates. 
Recovery time is here defined as the time between the discovery observation and the follow-up observation following the NEOCP posting of the first tracklet. 
Figure ~\ref{delay_followup1} further confirms that delayed submission is correlated with delayed recovery time.

It is clear that submission and recovery delays vary for different surveys. For instance, Mt. Lemmon and Catalina sites, operated by the same pipeline, are able to submit and recover most of their NEOCP candidates within 2 hours. Other surveys display much slower turnout, e.g., Pan-STARRS reported NEOCP candidates on average 10 hours after the initial observation and follow-up was reported on average 20 hours later. 
DECAM, SST and NEOWISE also display long submission and recovery delays.

\begin{figure}[ht!]
\center
\includegraphics[width=0.49\textwidth]{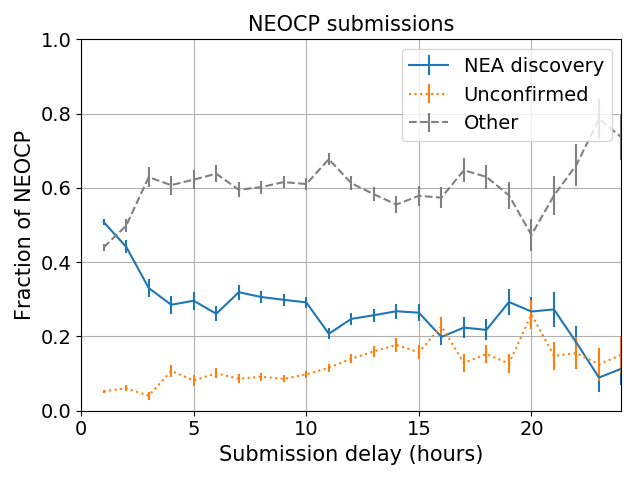}
\includegraphics[width=0.49\textwidth]{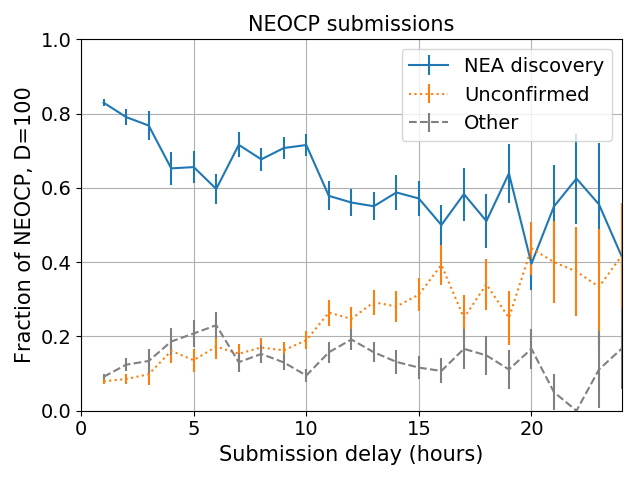}
\caption{Fraction of discoveries and unconfirmed NEOCP candidates as a mean submission delay for all (left) and $D_2=100$ (right) NEOCP candidates. NEO discovery rate drops and loss rate increases with increasing submission delay.}
\label{delay_fraction}
\end{figure}

We investigated how the submission delay translates to the properties of confirmed and unconfirmed NEA candidates. Figure~\ref{delay_fraction} demonstrates, that as a function of submission delay, the ratio of unconfirmed NEA candidates grows and the ratio of NEA discovery drops. This pattern is even stronger among $D_2=100$ candidates, that we assume are predominantly NEA. For instance, if an NEOCP candidate with $D_2=100$ is submitted within 3 hours, there is only a $10\%$ chance of it remaining unconfirmed.
In contrast, a 10-hour time delay (which is the mean time delay for Pan-STARRS) doubles the ratio of unconfirmed NEA candidates to $20\%$.
We note that $25\%$ of the unconfirmed NEA candidates from Pan-STARRS were submitted more than 15 hours after the discovery and for those the loss rate is about $30\%$. If Pan-STARRS could decrease the submission delay to 3 hours in the studied time period, their unconfirmed contribution of $D_2=100$ candidates would be $50\%$ smaller and based on Table~\ref{Htable}, Pan-STARRS unconfirmed NEA candidates would be by 173 lower.

There is an obvious correlation between the speed with which surveys submit their data (Figure \ref{delay_followup}) and the fraction of NEA candidates unconfirmed by the surveys (Figure~\ref{class_fraction}). 
For instance the two most productive telescopes, Mt. Lemmon and Pan-STARRS, have similar limiting magnitudes (Figure~\ref{magnitude}), but Pan-STARRS loses twice as many NEA candidates as Mt. Lemmon as a ratio of their own NEOCP submissions (Figure~\ref{class_fraction}), consistent with the submission delays for Pan-STARRS and Mt. Lemmon in Figure~\ref{delay_fraction}).

\subsection{Apparent rate of motion}
\begin{figure}[ht!]
\center
\includegraphics[width=0.49\textwidth]{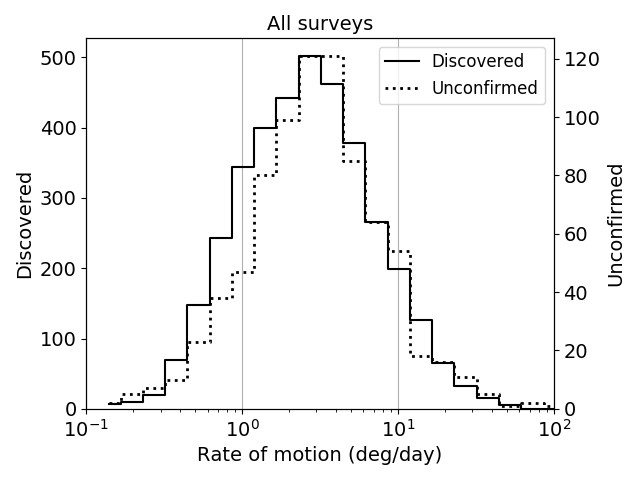}
\includegraphics[width=0.49\textwidth]{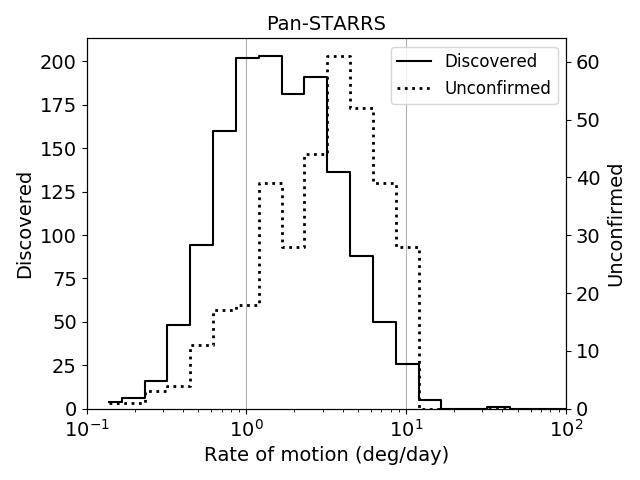}
\includegraphics[width=0.49\textwidth]{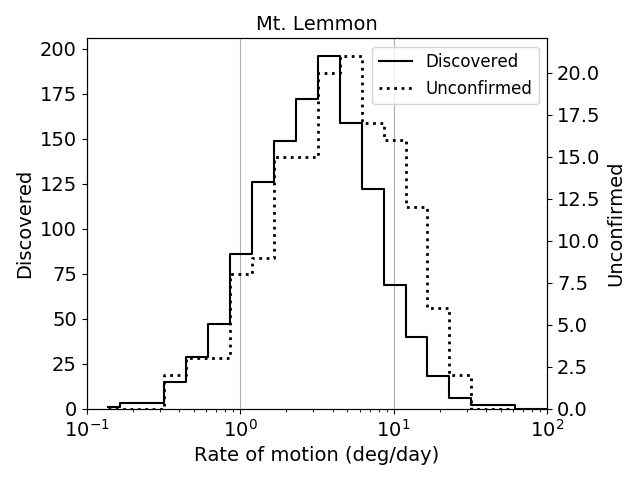}
\includegraphics[width=0.49\textwidth]{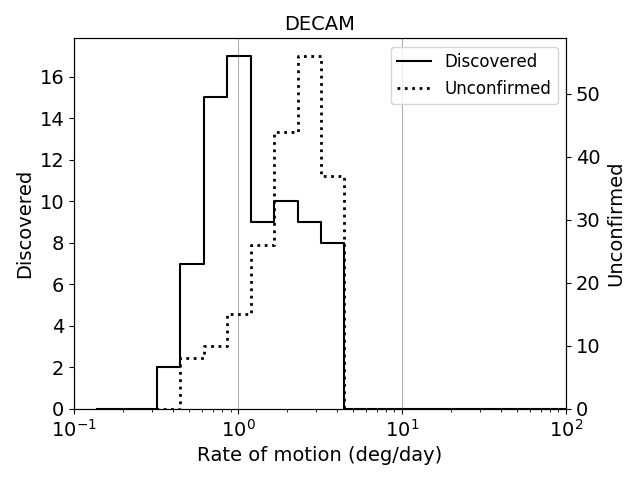}
\caption{Rate of motion of discovered and unconfirmed NEA candidates by surveys, $D_2=100$.
Unconfirmed NEA candidates have larger sky-plane motion than discovered NEAs.}
\label{rates}
\end{figure}

\begin{figure}[ht!]
\center
\includegraphics[width=0.7\textwidth]{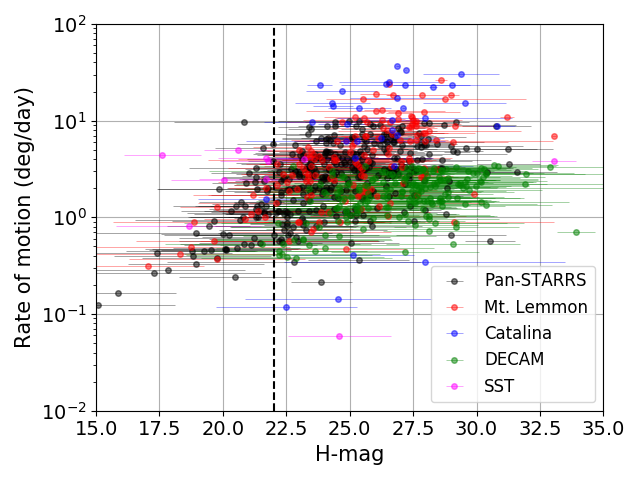}
\caption{Absolute magnitude vs. rate-of-motion (right) of unconfirmed NEOCP candidates with $D_2=100$. Largest unconfirmed NEA candidates are moving at the slower rate.}
\label{HvsRate}
\end{figure}

\begin{figure}[ht!]
\center
\includegraphics[width=0.7\textwidth]{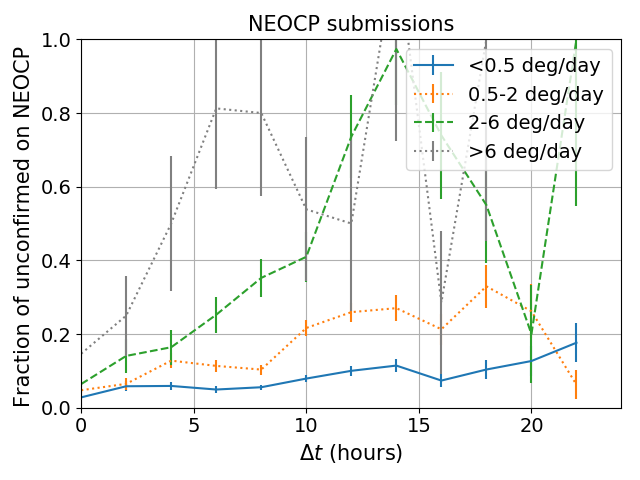}
\caption{Fraction of unconfirmed NEOCP candidates as a function of submission delay, divided into rate of motion bins. Increasing rate of motion causes larger loss rate at the same time delay. Fast NEOCP candidates require immediate submission.}
\label{rates1}
\end{figure}

A high apparent rate of motion increases the astrometric uncertainty along the apparent motion vector and also decreases the reported magnitude when performing either traditional Point-Spread-Function or aperture photometry \citep{Veres12,Fraser16}. 
\citet{Veres17} demonstrated that rate of motion correlates with the along-track residuals.

Figure~\ref{rates} shows that unconfirmed NEA candidates are  faster than the discovered ones for Pan-STARRS, Mt. Lemmon and DECAM. The mean rate of unconfirmed candidates is $4.2\,deg/day$ and $3.5\,deg/day$ for discovered NEAs. The figure shows that surveys can detect fast moving NEAs only up to a certain velocity limit, e.g. Pan-STARRS up to $10\,deg/day$, DECAM $3\,deg/day$. Mt. Lemmon is able to report faster NEA candidates.  Pan-STARRS displayed a larger difference in rate of motion of unconfirmed and discovered NEAs than Mt. Lemmon, also the submission delay was much larger for Pan-STARRS. This is because Mt. Lemmon has its own follow-up facilities and focuses on rapid follow-up. Fast (therefore nearby) NEA candidates are harder to track as the rate of motion increases, and the smallest unconfirmed NEA candidates are the fastest ones. 

On the other hand, the large unconfirmed NEA candidates with $H<22$ move at a slower pace (Figure~\ref{HvsRate}), primarily because larger objects are typically discovered farther away and their apparent motion is therefore lower. 
Additionally, fast moving NEA candidates are also prone to loss when they are submitted late. 
Figure~\ref{rates1} demonstrates that the fraction of unconfirmed candidates is larger for faster objects and depends on submission delay.

\subsection{Sky-plane location}

\begin{figure}[ht!]
\center
\includegraphics[width=1.0\textwidth]{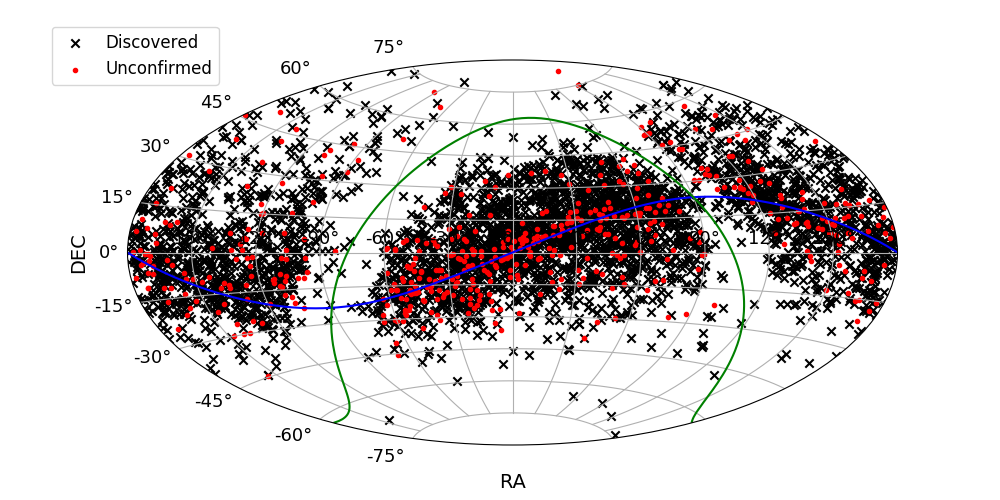}
\caption{Equatorial coordinates of discovered (black crosses) and unconfirmed NEA candidates (red dots) with $D_2=100$, the ecliptic is depicted in blue, the galactic plane in green. Unconfirmed and discovered NEAs do not cluster in any preferred sky-plane location and are similarly distributed.}
\label{sky1}
\end{figure}

NEAs are mostly discovered close to the ecliptic and near  opposition (Figure~\ref{sky1}). 
Although the spread of NEAs along the ecliptic is wider than that of main-belt asteroids, the concentration is still obvious. On the other hand, surveys in both the southern and northern hemispheres primarily focus on near-ecliptic fields. 
The map also shows that surveys avoid the galactic plane due to the large background star density, which makes it difficult to identify minor planets.

\begin{figure}[ht!]
\center
\includegraphics[width=0.48\textwidth]{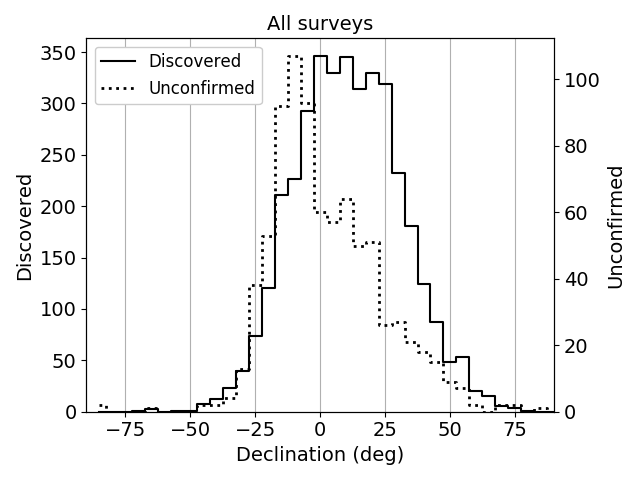}
\includegraphics[width=0.48\textwidth]{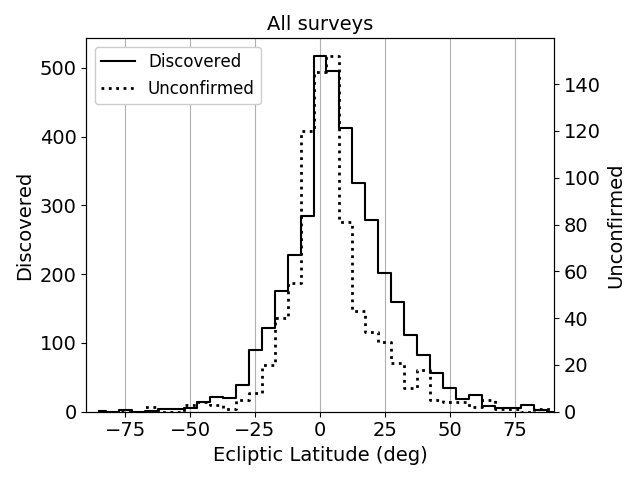}
\includegraphics[width=0.48\textwidth]{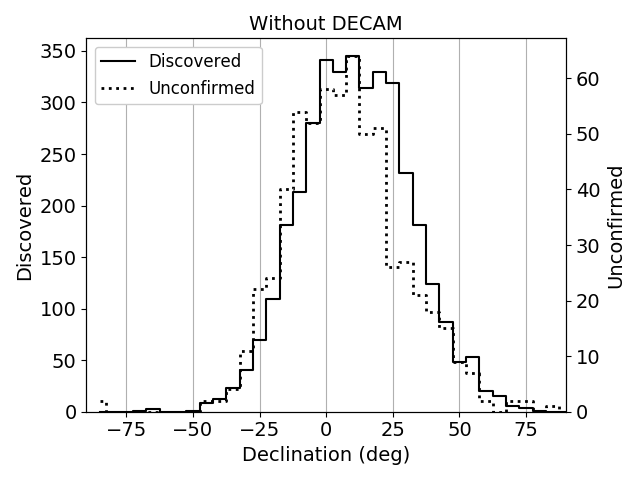}
\includegraphics[width=0.48\textwidth]{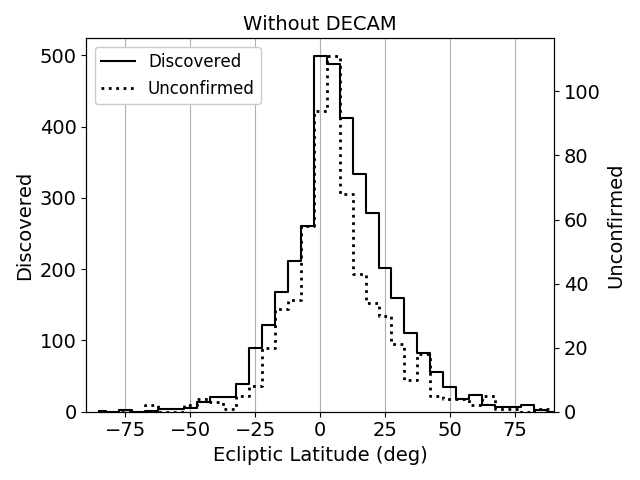}
\caption{Histograms of unconfirmed and discovered ($D_2=100$) NEA candidates in declination and ecliptic latitude  by surveys. The DECAM survey is the source of the excess of southern hemisphere unconfirmed NEAs. The bottom plots demonstrate that without DECAM data there is no difference between unconfirmed and discovered NEAs in either ecliptic or equatorial coordinates.}
\label{on_sky}
\end{figure}

At first glance, NEAs seem more likely to be unconfirmed south of the equator (Figure~\ref{on_sky}).
This is primarily due to the presence of the DECAM survey in the southern hemisphere, which produces a large amount of unconfirmed candidates: the bottom panels of Figure~\ref{on_sky} demonstrate that, when DECAM data are removed, the ratio of unconfirmed-to-discovered becomes essentially the same in both northern and southern hemispheres.

One might naively expect that the ratio between northern and southern hemisphere NEOCP candidates would be larger than one, based on the presence of more dedicated NEA surveys north of equator. 
However, Figure~\ref{on_sky} shows that there is no obvious evidence that unconfirmed NEA candidates are located differently on the sky than are the discovered objects.

\subsection{Seasonal changes}

\begin{figure}[ht!]
\center
\includegraphics[width=0.49\textwidth]{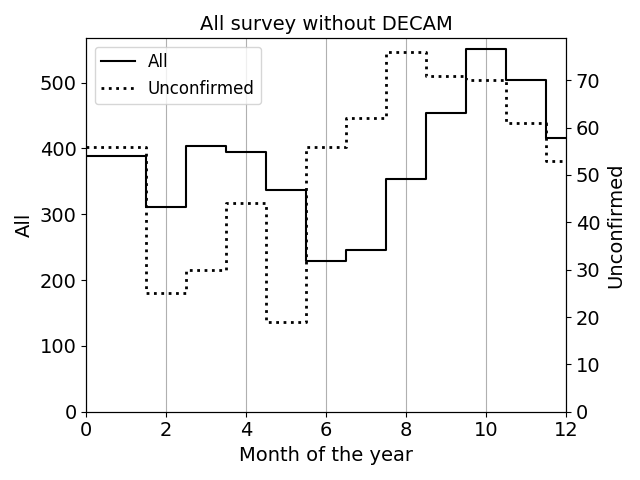}
\includegraphics[width=0.49\textwidth]{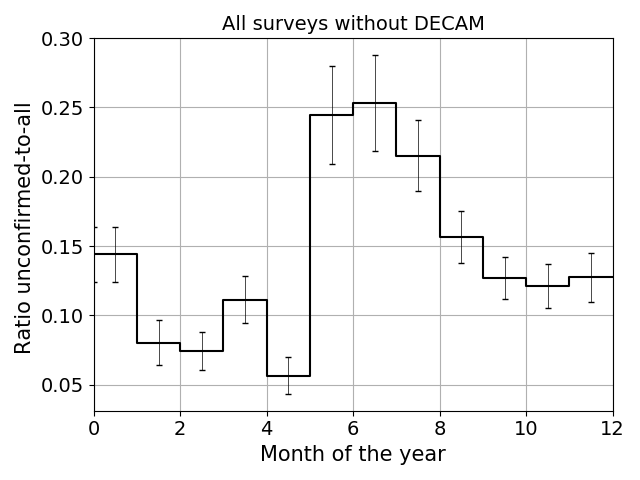}

\includegraphics[width=0.49\textwidth]{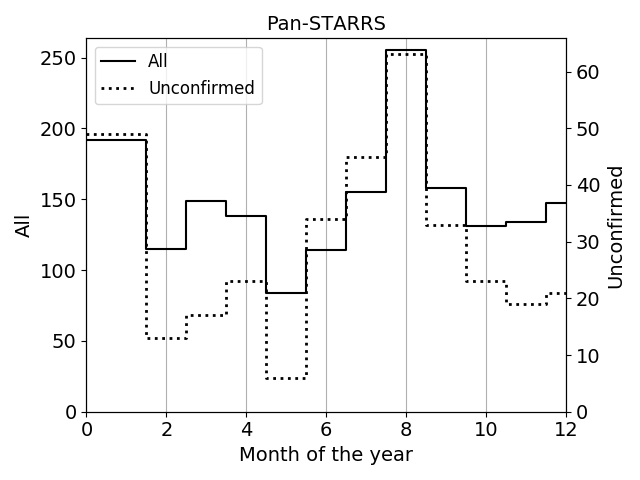}
\includegraphics[width=0.49\textwidth]{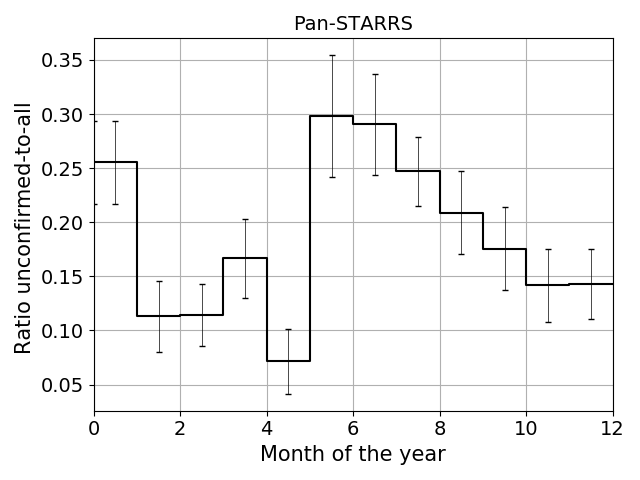}

\includegraphics[width=0.49\textwidth]{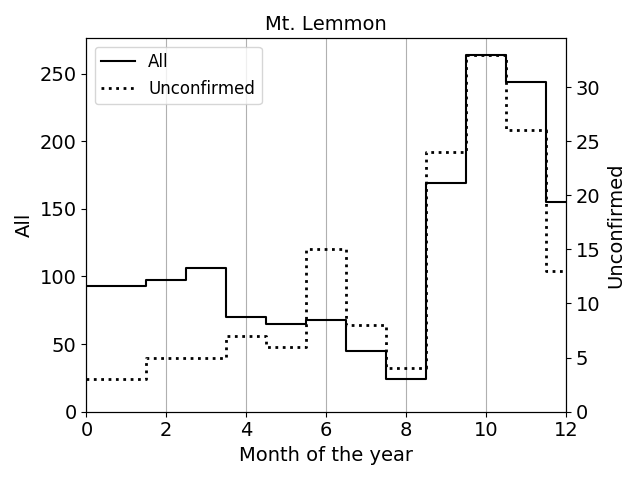}
\includegraphics[width=0.49\textwidth]{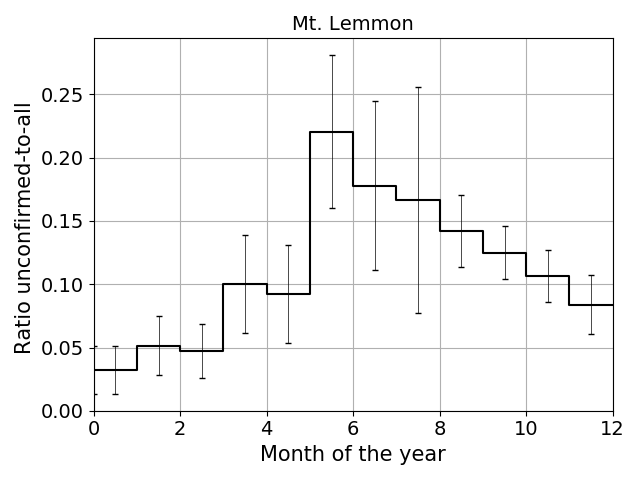}

\caption{{\bf Left:}  Number of discovered and unconfirmed NEA candidates by month of the year by surveys,$D_2=100$. Seasonal variations are visible and site-specific. {\bf Right:} Fraction of unconfirmed NEA
 candidates with respect to the sum of unconfirmed and discovered by month of the year.}
\label{month}
\end{figure}

\begin{figure}[ht!]
\center
\includegraphics[width=0.49\textwidth]{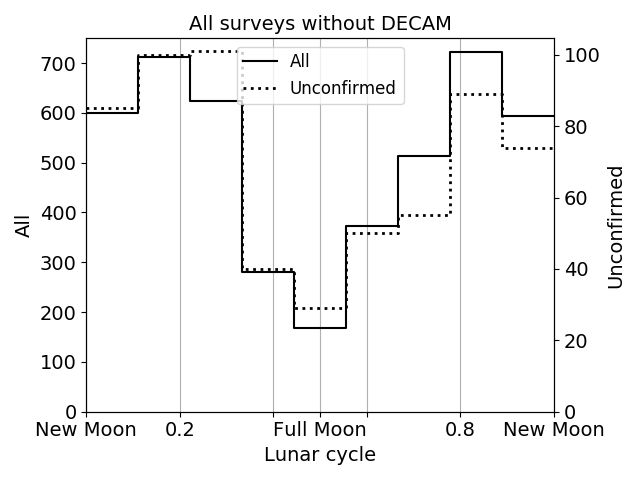}
\includegraphics[width=0.49\textwidth]{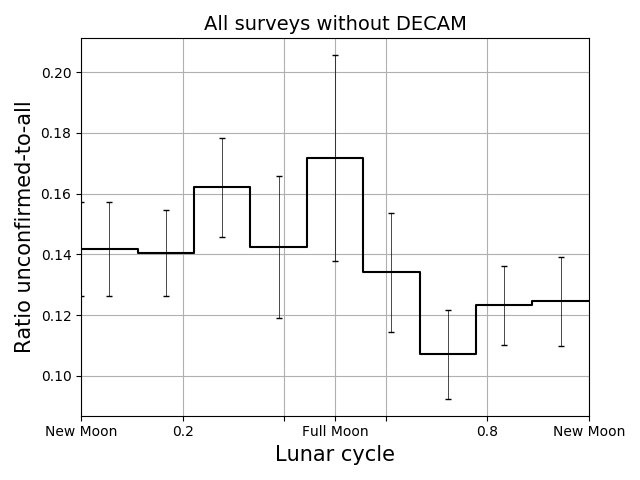}

\includegraphics[width=0.49\textwidth]{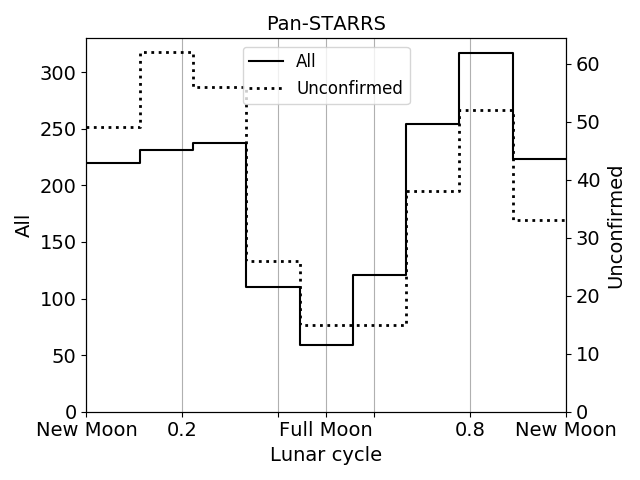}
\includegraphics[width=0.49\textwidth]{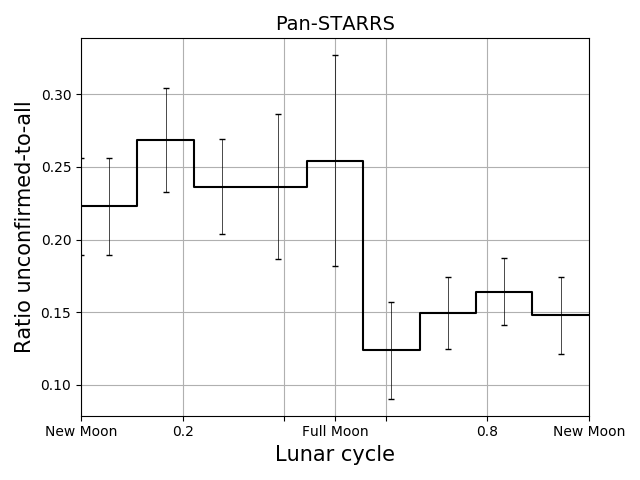}

\includegraphics[width=0.49\textwidth]{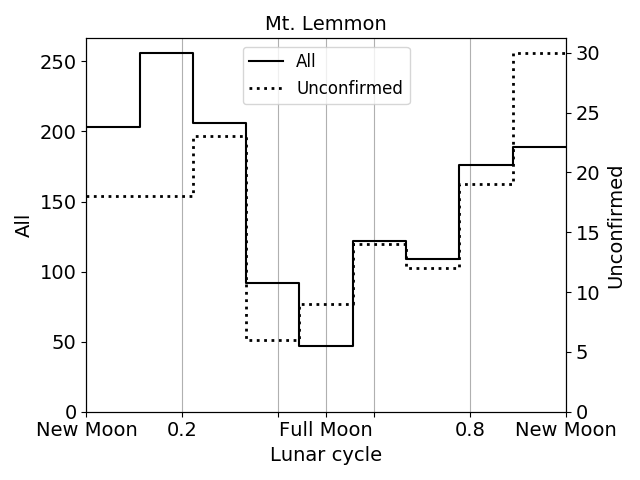}
\includegraphics[width=0.49\textwidth]{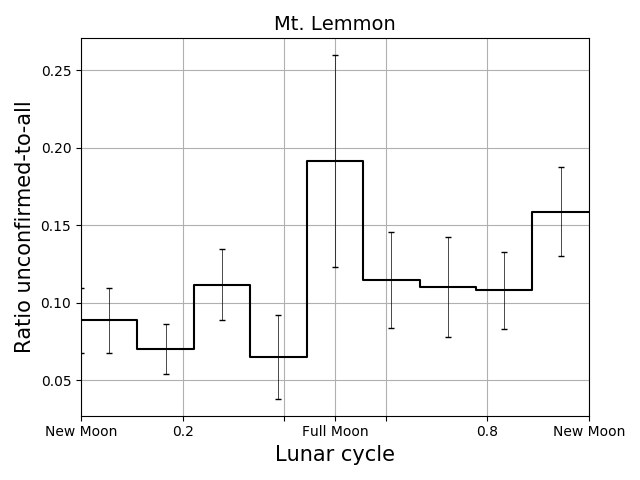}

\caption{{\bf Left:} Number of discovered and unconfirmed NEA candidates by phase of the Moon by surveys $D_2=100$. {\bf Right:} Fraction of unconfirmed NEA candidates with respect to the sum of unconfirmed and discovered NEOCP candidates by phase of the Moon. The phase of the Moon does not seem to play a significant role in the ratio of unconfirmed to discovered NEAs.}
\label{moon}
\end{figure}

The operation of ground-based telescopes is affected by atmospheric conditions, weather, and the Moon. For instance, surveys in Arizona (Catalina Sky Survey) are prone to limited operations due to monsoon season in the summer, while the winter season brings more cloud coverage to Hawaii (Pan-STARRS). The Moon is also a particular obstruction, increasing the sky background and decreasing the area of the sky available to survey. We studied Pan-STARRS, Mt. Lemmon and DECAM NEA candidates as a function of month of the year (Figure~\ref{month}) and lunar phase (Figure~\ref{moon}). DECAM contributed only in a few months and therefore we removed it from our analysis. 
The seasonal changes are strongly site-dependent.  Mt. Lemmon shows a correlation between unconfirmed and all NEA candidates during the year, with a surplus of unconfirmed NEA candidates in the summer months. Pan-STARRS unconfirmed NEA candidates display surplus in January and in the summer, the lowest ratio was from February to April. The summer discrepancy is due to a lack of follow-up from Arizona sites during the monsoon season. Telescopes in Arizona are also significantly contributing to the Pan-STARRS follow-up, therefore, the unconfirmed NEAs are more likely unconfirmed in the summer months in the combined data set as well.

Figure~\ref{moon} shows how the Moon and its phases affects the discovery and loss rates throughout the lunar cycle. 
Significantly fewer NEA candidates are reported around Full Moon and most NEA candidates are around the New Moon. 
The ratio of unconfirmed NEA candidates on the NEOCP is largest during the full Moon, which was expected, because a Full Moon makes large portion of the sky unobservable.  Pan-STARRS has proportionally more unconfirmed candidates right after New Moon than Mt. Lemmon. On the other hand, Pan-STARRS has low unconfirmed ratio after the Full Moon until the next New Moon, while Mt. Lemmon has a relatively large ratio at this time.

\clearpage

\section{Discussion}

NEO discovery and characterization are of great interest to NASA, the planetary defense community, and private industry.
Significant resources are being spent on telescopic surveys and NEA search. 
In this paper we showed that about $10\%$ of all NEOCP submissions are left without follow-up and remain unconfirmed (these represent a mixture of NEA and non-NEA candidates). 
We have estimated the number of observed but unconfirmed NEAs during the 2013--2016 period to be $926\pm 50$, approximately $18\%$ of the total number of NEAs successfully discovered during that period.

We find a number of causes for observed NEA candidates that are not being followed-up:
\begin{enumerate}
\item Submission delay affects the loss rate of NEOCP candidates. Increasing submission delays from 2 to 10 hours doubles the rate of unconfirmed NEA candidates. Timely submission of the collected astrometry to the MPC should be a clear priority of NEO surveys.
\item Fast rate of motion was a prominent feature of unconfirmed NEAs. Combination of fast rate of motion and delayed submission clearly showed increased loss rate of NEOCP candidates. Improved astrometry of fast moving targets and early follow-up decreases the number of unconfirmed and fast NEAs.
\item Unconfirmed NEA candidates have larger absolute magnitudes than discovered NEAs in the same time period. We find that only $11\%$ of unconfirmed NEA candidates had $H<22$.  DECAM contributed the majority of the smallest objects. Smaller NEAs are discovered when they are closer and moving fast, so their uncertainty grows more rapidly.
\item Surveys with faint limiting magnitude, such as DECAM, have much larger loss rate than Pan-STARRS and Mt. Lemmon, particularly for faint discoveries. A large telescope without secured follow-up is inefficient at turning its NEA candidates into NEAs discoveries. 

\item Seasonal weather changes and Moon affect the number of reported NEOCP candidates and should be taken into account when prioritizing the NEOCP targets, even though these are not a dominant reason of NEA candidate loss.

\end{enumerate}

We have demonstrated that some unconfirmed NEA candidates are later linked. The astrometry for any observed and submitted tracklet (NEA or otherwise) is stored in ITF at the MPC and is available to the community\footnote{\url{https://www.minorplanetcenter.net/iau/ECS/MPCAT-OBS/MPCAT-OBS.html}}. If the objects are observed in the future, either via serendipitous or  targeted follow-up, it is possible to link the ITF observations to the new observations, providing a means for those that once were unconfirmed to become found. 
However, because the rate of NEA discoveries from ITF linking is low, it is of significant benefit to reduce the number of NEOCP candidates that remain unconfirmed. 
 
The NEOCP follow-up would benefit from early submission of discovered and fast-moving NEA candidates. A global coordinated network of follow-up telescopes that cover the entire longitudinal and latitudinal range and desired magnitude depth, could improve the weather blackout seasons like late summer in Arizona and decrease the recovery time of posted NEOCP candidates.

Our results demonstrate that, for the successful detection of NEOs, it is critically important that surveys rapidly submit observations to facilitate the prompt follow-up of candidate detections. 
With new facilities starting their operations (e.g. Pan-STARRS2 \citep{Lilly} and the Zwicky Transient Facility \citep{Bellm2014}), and as the Large Synoptic Survey Telescope \citep[LSST,][]{Ivezic} builds towards full operations in 2022, we expect to see at least an order of magnitude increase in the rate of NEO discoveries \citep{Jones16,Jones18}. 
Without some form of prompt follow-up (either self-follow-up by survey design or separate facilities), our results suggest that the number of unconfirmed NEO candidates may be large, as we highlight is currently the case for the DECAM survey.
We advocate for 
(i) the rapid reporting of data to the MPC, 
(ii) the development of highly efficient linking tools, 
and 
(iii) the redoubling of community efforts to plan for the efficient follow-up of current and future NEO survey detections as they push to ever-fainter detection limits.

\clearpage
 
 \section*{Acknowledgement}
We are thankful to the reviewer for their thorough review and helpful comments.
D. Farnocchia conducted this research at the Jet propulsion Laboratory, California Institute of Technology, under a contract with NASA. 
The authors at the MPC gratefully acknowledge the NASA/University of Maryland MPC Operations grant 46039-Z6110001, the NASA Solar System Observations grants NNX16AD69G and NNX17AG87G, 
as well as support from the Smithsonian Scholarly Studies program.

\appendix
\section{Auxiliary Tables}

\begin{table*}[ht!]
\small
\begin{center}
\caption{NEOCP rules for issuing MPEC for NEA discovery.}
\vspace{1em}
\begin{tabular}{l|c|c|c}
\hline
Type     & H    &Arc (days)  & N (tracklets)\\
\hline
MPEC  &     any  &  0.0&     10\\
MPEC &      $>26$   & 0.5  &    3\\
MPEC &      $23-26$ &   1.2  &    3\\
MPEC &      $20-23$   & 1.9   &   4\\
MPEC  &    $<20.0$   & 2.6  &    4\\
non-MPEC  & any   & 0.833 &   3\\
\hline
\end{tabular}
\label{neocp}
\end{center}
\end{table*}

\begin{table*}[ht!]
\small
\begin{center}
\caption{Transformation of the reported magnitude band to Johnson V-band used by the MPC.}
\begin{tabular}{c|c||c|c}
\hline
Band & V correction& Band & V mag correction\\
\hline
&-0.8&z&0.26\\
U&-1.3&I&0.8\\
B&-0.8&J&1.2\\
g&-0.35&w&-0.13\\
V&0&y&0.32\\
r&0.14&L&0.2\\
R&0.4&H&1.4\\
C&0.4&K&1.7\\
W&0.4&Y&0.7\\
i&0.32&G&0.28\\
\hline
\end{tabular}
\label{vband}
\end{center}
\end{table*}

\end{document}